\documentclass[twocolumn,showpacs,aps,prl,superscriptaddress]{revtex4}

\usepackage{graphicx}
\usepackage{dcolumn}
\usepackage{amsmath}
\usepackage{epsfig}
\usepackage{bm}

\input pubboard/babarsym
\newcommand{\bma}[1]{\boldmath{$#1$}}

\providecommand{\goto}{\rightarrow}
\providecommand{\calB}{\ensuremath{{\cal B}}}

\def\dbline{\noalign{\vskip 0.15truecm\hrule}\noalign{\vskip 2pt}\noalign{\hrule\vskip 0.15truecm}}

\def\tbline{\noalign{\vskip 0.05truecm\hrule\vskip 0.05truecm}}

\newcommand{\respi}{\ensuremath{B^+\ra R\pi^+}}
\newcommand{\resKp}{\ensuremath{B^+\ra R K^+}}

\providecommand{\etapr}{\ensuremath{\eta^\prime}}

\newcommand{\fetapeppKp}{\ensuremath{\etapr_{\eta\pi\pi} K^+}}
\newcommand{\fetaprgKp}{\ensuremath{\etapr_{\rho\gamma} K^+}}
\newcommand{\fetapeppKz}{\ensuremath{\etapr_{\eta\pi\pi} K^0}} 
\newcommand{\fetaprgKz}{\ensuremath{\etapr_{\rho\gamma} K^0}} 
\newcommand{\fetapepppp}{\ensuremath{\etapr_{\eta\pi\pi} \pi^+}}
\newcommand{\fetaprgpp}{\ensuremath{\etapr_{\rho\gamma} \pi^+}}

\newcommand{\etapK}{\ensuremath{B\ra\etapr K}}

\newcommand{\fetapKp}{\ensuremath{\etapr K^+}}
\newcommand{\etapKp}{\ensuremath{B^+\ra\fetapKp}}
\newcommand{\BetapKp}{\ensuremath{\calB(\etapKp)}}
\newcommand{\retapKp}{\ensuremath{70 \pm 8 \pm 5}}
\newcommand{\RetapKp}{\ensuremath{(\retapKp)\times 10^{-6}}}

\newcommand{\fetapKz}{\ensuremath{\etapr K^0}}
\newcommand{\etapKz}{\ensuremath{B^0\ra\fetapKz}} 
\newcommand{\BetapKz}{\ensuremath{\calB(\etapKz)}}
\newcommand{\retapKz}{\ensuremath{42^{+13}_{-11} \pm 4}}
\newcommand{\RetapKz}{\ensuremath{(\retapKz)\times 10^{-6}}}

\newcommand{\fetappp}{\ensuremath{\etapr \pi^+}}
\newcommand{\etappp}{\ensuremath{B^+\ra\fetappp}}
\newcommand{\Betappp}{\ensuremath{\calB(\etappp)}}
\newcommand{\retappp}{\ensuremath{ 5.4^{+3.5}_{-2.6} \pm 0.8}}

\newcommand{\uetappp}{\ensuremath{12}}
\newcommand{\Uetappp}{\ensuremath{\uetappp\times 10^{-6}}}

\newcommand{\fomegapi}{\ensuremath{\omega\pi^+}}
\newcommand{\omegapi}{\ensuremath{B^+\ra\fomegapi}}
\newcommand{\Bomegapi}{\ensuremath{\calB(B^+\ra\omega\pi^+)}}
\newcommand{\romegapi}{\ensuremath{6.6^{+2.1}_{-1.8}\pm 0.7}}
\newcommand{\Romegapi}{\ensuremath{(\romegapi)\times 10^{-6}}}

\newcommand{\fomegapiz}{\ensuremath{\omega\piz}}
\newcommand{\omegapiz}{\ensuremath{B^0\ra\fomegapiz}}
\newcommand{\Bomegapiz}{\ensuremath{\calB(\omegapiz)}}
\newcommand{\romegapiz}{\ensuremath{-0.3\pm1.1\pm 0.3}}

\newcommand{\uomegapiz}{\ensuremath{3}}
\newcommand{\Uomegapiz}{\ensuremath{\uomegapiz\times 10^{-6}}}

\newcommand{\fomegaKp}{\ensuremath{\omega K^+}}
\newcommand{\omegaKp}{\ensuremath{B^+\ra\fomegaKp}}
\newcommand{\BomegaKp}{\ensuremath{\calB(\omegaKp)}}
\newcommand{\romegaKp}{\ensuremath{1.4^{+1.3}_{-1.0}\pm 0.3}}

\newcommand{\uomegaKp}{\ensuremath{4}}
\newcommand{\UomegaKp}{\ensuremath{\uomegaKp\times 10^{-6}}}

\newcommand{\fomegaKz}{\ensuremath{\omega K^0}}
\newcommand{\omegaKz}{\ensuremath{B^0\ra\fomegaKz}}
\newcommand{\BomegaKz}{\ensuremath{\calB(\omegaKz)}}
\newcommand{\romegaKz}{\ensuremath{6.4^{+3.6}_{-2.8}\pm 0.8}}

\newcommand{\uomegaKz}{\ensuremath{13}}
\newcommand{\UomegaKz}{\ensuremath{\uomegaKz\times 10^{-6}}}

\newcommand{\etagg}{\ensuremath{\eta\ra\gaga}}

\newcommand{\omegathrpi}{\ensuremath{\omega\ra\pi^+\pi^-\pi^0}}

\newcommand{\etaprd}{\ensuremath{\etapr\ra\eta\pi^+\pi^-}}

\newcommand{\DE}{\ensuremath{\Delta E}}
\newcommand{\mb}{\ensuremath{\mec}}
\newcommand{\xf}{\ensuremath{{\cal F}}}
\newcommand{\hel}{\ensuremath{{\cal H}}}
\newcommand{\costhr}{\ensuremath{\cos\theta_{\rm T}}}
\providecommand{\KS}{\ensuremath{K_S^0}}
\newcommand{\kzs}{\KS}

\providecommand{\UfourS}{\ensuremath{\Upsilon(4S)}}

\providecommand{\n}{\mbox{\phantom{4}}}

\newcommand{\Cherenkov}{Cherenkov}

\def\Bz{\mbox{$\overline {B^{0}}$}}

\def\BB{\mbox{$\B^{+}\  \B^{-}$}}

\def\Bz{\mbox{$B^0\ $}}
\def\Bzb{\mbox{$\overline{B}^0\ $}}
\def\beq{\begin{equation}}
\def\eeq{\end{equation}}
\def\bef{\begin{figure}}
\def\edf{\end{figure}}
\def\ben{\begin{enumerate}}
\def\een{\end{enumerate}}
\def\bear{\begin{array}}
\def\enar{\end{array}}
\def\beqa{\begin{eqnarray}}
\def\eeqa{\end{eqnarray}}

\def\to{\mbox{$\rightarrow$}}

\def\gev{\mbox{${\mathrm{GeV}}$}}
\def\gevc{\mbox{${\mathrm{GeV}}/c$}}
\def\gevcc{\mbox{${\mathrm{GeV}}/c^2$}}

\def\mevcc{\mbox{${\mathrm{MeV}}/c^2$}}

\def\FourS{\mbox{$\Upsilon{\mathrm( 4S)}$}}

\def\Bz{\mbox{${B^{0}}$}}
\def\Bzb{\mbox{$\overline B^{0}$}}
\def\BB{\mbox{$B\overline B$}}
\def\BpBm{\mbox{$B^{+} B^{-}$}}

\def\piz{\mbox{${\pi^{0}}$}}
\def\pip{\mbox{${\pi^{+}}$}}
\def\pim{\mbox{${\pi^{-}}$}}

\def\pimp{\mbox{${\pi^{\mp}}$}}

\def\Kp{\mbox{${K^{+}}$}}
\def\Km{\mbox{${K^{-}}$}}

\def\DE{\mbox{${\Delta E}$}}

\newcommand{\BABARPubYear}    {01}
\newcommand{\BABARPubNumber}  {15}

\newcommand{\SLACPubNumber} {8956}

\def\figurebox#1#2#3{%
    \def\arg{#3}%
    \ifx\arg\empty
    {\hfill\vbox{\hsize#2\hrule\hbox to #2{\vrule\hfill\vbox to #1{\hsize#2\vfill}\vrule}\hrule}\hfill}%
    \else
    {\hfill\epsfbox{#3}\hfill}%
    \fi}

\begin{document}

\preprint{\babar-PUB-\BABARPubYear/\BABARPubNumber} 
\preprint{SLAC-PUB-\SLACPubNumber} 

\begin{flushleft}
\babar-PUB-\BABARPubYear/\BABARPubNumber\\
SLAC-PUB-\SLACPubNumber\\
\end{flushleft}

\title{
{\large \bf
  Measurements of the branching fractions of exclusive charmless
\boldmath{$B$} meson decays with \boldmath{\etapr}\ or 
\boldmath{$\omega$}\ mesons
} 
}

%
\author{B.~Aubert}
\author{D.~Boutigny}
\author{J.-M.~Gaillard}
\author{A.~Hicheur}
\author{Y.~Karyotakis}
\author{J.~P.~Lees}
\author{P.~Robbe}
\author{V.~Tisserand}
\affiliation{Laboratoire de Physique des Particules, F-74941 Annecy-le-Vieux, France }
\author{A.~Palano}
\affiliation{Universit\`a di Bari, Dipartimento di Fisica and INFN, I-70126 Bari, Italy }
\author{G.~P.~Chen}
\author{J.~C.~Chen}
\author{N.~D.~Qi}
\author{G.~Rong}
\author{P.~Wang}
\author{Y.~S.~Zhu}
\affiliation{Institute of High Energy Physics, Beijing 100039, China }
% ...
\collaboration{The \babar\ Collaboration}
\noaffiliation

\date{October 4, 2001}

\begin{abstract}
We present the results of searches for $B$ decays to
charmless two-body final states containing \etapr\ or $\omega$ mesons,
based on 20.7 fb$^{-1}$ of data collected with the \babar\ detector.
We find the branching fractions
 $\BetapKp = \RetapKp$, 
 $\BetapKz = \RetapKz$, and
 $\Bomegapi = \Romegapi$
where the first error quoted is statistical and the second systematic.
We give measurements of four additional modes for which the 90\%
confidence level upper limits are
 $\BomegaKz < \UomegaKz$,
 $\Betappp < \Uetappp$,
 $\BomegaKp < \UomegaKp$, and
 $\Bomegapiz < \Uomegapiz$.
\end{abstract}


\maketitle
%
%
We report results of searches for $B$ decays to the charmless
two-body final states 
\cite{bib:conjugate}\ \omegapi, \omegaKp, \etappp, \etapKp, \omegaKz,
\omegapiz, and  \etapKz.  These processes 
are manifestations of penguin or suppressed tree amplitudes proportional
to small couplings in hadronic flavor mixing (CKM matrix
\cite{CKM}).  Because of the absence of CKM favored $b\goto c$ amplitudes these
decays are particularly sensitive to potentially new contributions from
interference 
effects and virtual particles in loops.  Previous measurements
\cite{CLEOetapr} yielded an unexpectedly large rate for \etapK,
motivating a number of new theoretical ideas.  
The precise measurement of these and
additional rare $B$ decay modes will enable a better understanding of the
underlying decay mechanism, including the possible contribution of
physics beyond the standard model. This in turn will contribute to the
measurement of fundamental parameters, including the $CP$-violating CKM
phases.

%
%
The data were collected with the \babar\ detector~\cite{BABARNIM}
at the PEP-II asymmetric $e^+e^-$ collider~\cite{pep}
located at the Stanford Linear Accelerator Center.
The results presented in this paper are based on data taken
in the 1999--2000 run. An integrated
luminosity of 20.7~fb$^{-1}$, corresponding to 
22.7 million \BB\ pairs, was recorded at the $\Upsilon (4S)$
resonance
(``on-resonance'', $10.58\ \gev$), with an additional 2.6~fb$^{-1}$ about 40~MeV below
this energy (``off-resonance'') for the study of continuum backgrounds. 

The asymmetric beam configuration in the laboratory frame
provides a boost to the $\Upsilon(4S)$
increasing the momentum range of the $B$-meson decay products
up to $4.3\ \gevc$.
Charged particles are detected and their momenta measured
by a combination of a silicon vertex tracker (SVT), consisting 
of five layers of double-sided detectors, and a 40-layer central drift chamber 
(DCH), both operating in the 1.5~T magnetic field of a solenoid. 
Photons and electrons are detected by a CsI(Tl) electromagnetic calorimeter
(EMC), which 
provides excellent angular and energy resolution with high efficiency for 
energies above 20~MeV~\cite{BABARNIM}.

Charged particle identification (PID) is provided by the average 
energy loss ($dE/dx$) in the tracking devices  and
by a unique, internally reflecting ring imaging 
\Cherenkov\ detector (DIRC) covering the central region. 
A \Cherenkov\ angle $K$--$\pi$ separation of better than 4 standard
deviations ($\sigma$) is 
achieved for tracks below $3\ \gevc$ momentum, decreasing to 
2.5$\sigma$ at the highest momenta in the final states considered here
\cite{KpiPRL}.  
Electrons are identified with the use of the EMC.

%
%
We reconstruct a $B$ meson candidate by combining an $\omega$ or
$\eta^\prime$ candidate with a charged track, $\piz\goto\gamma\gamma$,
or $\kzs\goto\pi^+\pi^-$.  The resonance decays $R$ we reconstruct are
\omegathrpi, \etaprd\ 
($\etapr_{\eta\pi\pi}$), or \etaprrg\ ($\etapr_{\rho\gamma}$), with
\etagg\ and $\rho^0\ra\pi^+\pi^-$.  These
modes are kinematically distinct from the dominant $B$ decays to heavier
charmed states.  Backgrounds come primarily from combinatorics among
continuum events in which a light quark pair is produced instead of an
\UfourS. 

Monte Carlo (MC) simulations \cite{geant}\ of the target decay modes and
of continuum background are used to establish the event selection criteria.
The selection is designed to achieve high 
efficiency and retain sidebands sufficient to characterize the
background for subsequent fitting.
Photons must satisfy $E_\gamma>$ 50 (100) MeV for
\piz\ ($\eta$) candidates.  For \etaprrg\ candidates from \etapKp\ and
\etappp\ the requirement is $E_\gamma>$ 200 MeV, while for \etapKz\ it
is looser ($E_\gamma>$ 100 MeV) because of smaller combinatoric
background.

We select $\omega$, $\eta^\prime$, $\eta$, and $\rho$ candidates 
with the following requirements on the invariant masses in \mevcc\ of
their final 
states: $735 < m(\pi^+\pi^-\pi^0) < 830$, $930 <
m(\eta\pi^+\pi^-) <990$, $900 < m(\rho\gamma) <1000$, $490 < m(\gamma\gamma) < 600$, and $500 < m(\pi^+\pi^-) <995$.  For
\piz\ and \kzs\ candidates we require 
$120 < m(\gamma\gamma) < 150$ and $488 < m(\pi^+\pi^-) < 508$.  

Tracks in $\omega$, \etapr, or $\rho$ candidates must have DIRC, $dE/dx$, and EMC
responses consistent with pions.  For charged $B$ decays, the $B$ primary
track must have an associated DIRC \Cherenkov\ angle within 3.5$\sigma$
of the expected value for a kaon or pion.  For modes with \kzs\ the
three-dimensional flight distance from the production point must exceed
2 mm, and the angle between the flight and momentum vectors projected
perpendicular to the beam must be less than 40 mrad.

A $B$ meson candidate is characterized by two kinematic observables.
The minimally correlated pair we use are the energy constrained mass
\mb\ and energy difference \DE.  In the \UfourS\ frame the $B$
meson energy $E^*$ equals the beam energy $E^*_{\rm beam}$.  A kinematic
fit of the measured candidate four momentum in this frame with the
constraint $E^* =
E^*_{\rm beam}$ yields \mb, while $\DE\equiv E^* - E^*_{\rm beam}$ measures the
consistency of this constraint.
We require
$|\DE|\le0.2$ GeV, and $\mb\ge5.2\ \gevcc$.  The resolutions on these
quantities are mode dependent but average about 30 MeV and $2.8\ \mevcc$,
respectively. 

To discriminate against tau-pair and two-photon background we require the
event to contain at least five charged tracks.  To reject continuum
background we make use of the angle $\theta_T$ between the thrust axes
of the $B$ candidate and the rest of the tracks and neutral clusters in
the event, calculated in the center-of-mass frame.  The distribution of
$\cos{\theta_T}$ is sharply peaked near 
$\pm1$ for combinations drawn from jetlike $q\bar q$ pairs, and nearly
uniform for the isotropic $B$ meson decays.

%
%
The yields are obtained from extended unbinned maximum likelihood (ML)
fits, with two variants specified in the following paragraphs.  The
first (ML1), which provides our results for all 
modes except 
\omegapiz, uses several uncorrelated variables for the
kinematics of the $B$
decay chain and a Fisher discriminant for
the production and energy flow.  
The second (ML2) is applied to all channels with an $\omega$ 
meson; it uses \DE\ and the output of a neural network built from the
remaining inputs.  Comparisons for the $\omega\pi^+$, $\omega K^+$, and
$\omega K^0$ modes show that
the central values and errors for the yields obtained by the two
approaches are in very good agreement.  Simple cut-based analyses are
performed as checks for each final state.  Agreement of central values
is good in all cases, although, as expected, errors are larger than for
the ML analyses, particularly for modes having high background.

The ML1 fit method is applied to events satisfying
$|\cos{\theta_T}|\le0.9$.  The input observables are
\DE, \mb, the invariant mass $m_R$ of the intermediate resonance, the Fisher
discriminant
\xf, and, where relevant, the $\eta$ mass $m_\eta$,
the measured DIRC \Cherenkov\ angle for the $B$ primary track, and the
cosine \hel\ of the helicity angle, the angle in the 
$\omega$ rest frame between the normal to the $\omega$ decay plane and
the $B$ flight direction.  
The Fisher discriminant 
\cite{CLEO-fisher}\ combines eleven variables:  the angles with respect
to the beam axis in the \UfourS\
frame of the $B$ momentum and $B$ two-body decay axis,
and a nine bin 
representation of the energy flow about the $B$ decay axis.  

For the ML2 method we relax the preliminary
requirements to $100 < m(\gamma\gamma) < 160\ \mevcc$ and
$|\DE|<0.3$ GeV.
The neural network is constructed with the $B$ momentum $p^*$, a
$\chi^2$ for resonance masses, \hel, and variables representing energy
flow and angular distributions, including $\theta_T$.  

We use MC to estimate backgrounds from other $B$ decays, including final
states with and without charm.  For most of our modes we find
contributions that are negligible.  For the \etaprrg\ modes we account
for small cross feed contributions in the systematic error estimate.

The likelihood function for $N$ events is
\begin{displaymath}
 {\cal L} = \frac{e^{-(\sum n_j)}}{N!} \prod_{i=1}^N {\cal L}_i\,,\quad
{\cal L}_i = \sum_{j=1}^{m}n_j {\cal P}_j({\bf x}_i).
\label{eq:evtL}
\end{displaymath}
Here $n_j$ is the population size for species $j$ (e.g., signal,
background) and ${\cal P}_j({\bf x}_i)$ the corresponding probability
distribution function (PDF),
evaluated with the observables ${\bf x}_i$ of the $i$th event.  

For the fits of charged $B$ decays ${\cal L}_i$ becomes
\begin{eqnarray*}
{\cal L}_i &=& n_\pi {\cal P}_{\pi S}({\bf x}_i) + n_K {\cal P}_{KS}({\bf x}_i) + \\
   && n_C \left[ f_{KC} {\cal P}_{KC}({\bf x}_i) + (1-f_{KC}) {\cal P}_{\pi C}({\bf x}_i)\right]\,,
\end{eqnarray*}
where $n_\pi$($n_K$) is the number of \respi(\resKp) signal events,
$n_C$ is the number of continuum background events, and $f_{KC}$ is the
fraction of continuum background events for which the $B$ primary track is
identified as a kaon. These quantities are the free parameters of the ML
fit. The probabilities for the components are ${\cal P}_{\pi S}$ (${\cal
P}_{KS}$) for \respi(\resKp) signal and ${\cal P}_{\pi C}$ (${\cal
P}_{KC}$) for background where the primary track is a pion (kaon).
Since we measure the correlations among the observables in the data to
be small, we take each ${\cal P}_j$ to be a product of the PDFs for the
separate observables.  The analyses 
involving a \KS\ are treated identically except that there is only one
component of signal and of continuum background.

A second $B$ candidate satisfying the preliminary cuts occurs in about
10--20\% of the events.  In this case the ``best'' combination is selected
according to a $\chi^2$ quantity computed with \mb, $m_R$, $m_\eta$ (for
\etaprd\ modes), and the Fisher discriminant.

\newcommand{\psfile}[3][]{ 
  \begin{center}
    \setlength{\epsfxsize}{#3\linewidth}\leavevmode
    \def\noOpt{}\def\testit{#1}\ifx\testit\noOpt%
      \epsfbox{#2}%
    \else%
      \epsfbox[#1]{#2}%
    \fi
  \end{center} 
}

\begin{figure}[htbp]
 \psfile[62 154 540 494]{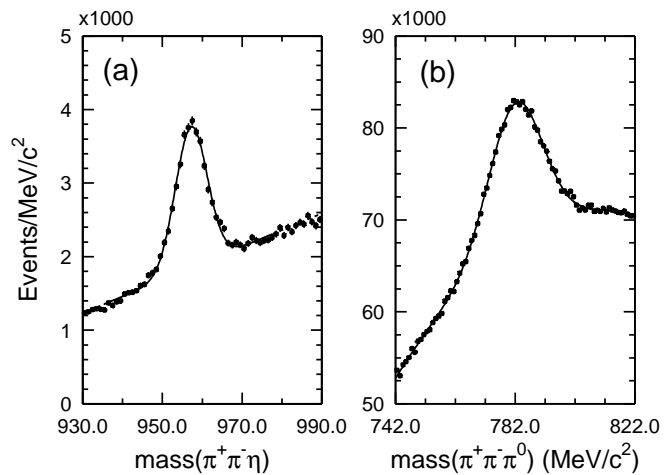}{1.0}
\vspace{-0.5cm}
 \caption{\label{fig:Rmass}
Invariant mass distributions for inclusive data
samples of candidates with \UfourS\ frame momentum greater than $2.3\
\gevc$ for (a) 
\etapr, with $520<m(\gamma\gamma)<575\ \mevcc$, and (b) $\omega$ 
candidates, with $120<m(\gamma\gamma)<150\ \mevcc$.  From the overlaid
fit curves the Gaussian peak widths are 4 and 
$10\ \mevcc$, respectively.
}
\end{figure}

We determine the PDFs for the likelihood fits from
simulation for the signal component, and from off-resonance and sideband
data for the continuum background.  Peaking distributions (signal
masses, \DE, \xf) are parameterized as Gaussians, with or without a
second Gaussian or asymmetric width as required to describe the
distributions.  Slowly varying distributions (combinatoric background
under mass 
or energy peaks, \hel, or \xf) have first or second order polynomial
shapes.  The combinatoric background in \mb\ is described by a phase
space motivated empirical function \cite{argus}.
Control samples of $B$ decays to charmed final states of similar topology
are used to verify the simulated resolutions in \DE\
and \mb.  Inclusive resonance production samples such as those shown in
Fig.\ \ref{fig:Rmass}\ are used
similarly for the relevant $B$ daughter mass spectra.

%
%
\providecommand{\corrEffB}{Corr. $\epsilon\times\prod\calB_i$ (\%)}
\providecommand{\signifSyst}{Signif. w syst. ($\sigma$)}
\providecommand{\bfemsix}{${\cal B}$}
\begin{table}[htbp]
\caption{
Signal event yield with statistical uncertainty, detection efficiency
$\epsilon$, daughter branching fractions that were forced to 100\% in our signal mode simulations, 
significance $S$ (defined in the text), and branching fraction result for each decay chain or mode, with the final (combined) result
given in bold type.  We show 90\% confidence upper limits in parentheses
where appropriate.
}
\label{t:results}
\begin{center}
\begin{tabular}{lccccccr}
\dbline
Mode		& Yield			& $\epsilon$	& $\prod\calB_i$& $S$		& \multicolumn{2}{c}{\bfemsix} 	\\
                &			& \%    	& \%		& $\sigma$	& \multicolumn{2}{c}{$10^{-6}$}	\\
\tbline                                                                 							
\n\fetapeppKp& $49.5^{+8.1}_{-7.3}$	& 20		& 17.4		& 15		& $63^{+10}_{-9}$	&	\\
\n\fetaprgKp	&$87.6^{+13.4}_{-12.5}$	& 18		& 29.5		& 11		& $80^{+12}_{-11}$	&	\\
\fetapKp	&			&		&		& 17		& \bma{\retapKp}	&	\\
\tbline                                                                 							
\n\fetapeppKz& $6.3^{+3.3}_{-2.5}$	& 16		& 6.0		& 4.7		& $28^{+15}_{-11}$	&	\\
\n\fetaprgKz	& $20.8^{+7.4}_{-6.5}$	& 16		& 10.1		& 4.2		& $61^{+22}_{-19}$	&	\\
\fetapKz	&			&		&		& 5.9		& \bma{\retapKz}	&	\\
\tbline                                                                 							
\n\fetapepppp& $5.7^{+3.8}_{-2.8}$	& 20		& 17.4		& 3.2		& $7.1^{+4.8}_{-3.5}$	&	\\
\n\fetaprgpp	& $-0.9^{+7.8}_{-6.2}$	& 19		& 29.5		& 0.1		& $-0.7^{+6.7}_{-5.3}$	&	\\
\fetappp	&			&		&	        & 2.8		& \bma{\retappp} 	&($<\uetappp$)	\\
\tbline                                                                 							
\fomegaKp	& $6.4^{+5.6}_{-4.4}$	& 22		& 88.8          & 1.3		& \bma{\romegaKp} 	&($<\uomegaKp$)	\\
\fomegaKz	& $8.1^{+4.6}_{-3.6}$	& 18		& 30.5		& 3.2		& \bma{\romegaKz} 	&($<\uomegaKz$)	\\
\fomegapi	& $27.6^{+8.8}_{-7.7}$	& 21		& 88.8		& 4.9		& \bma{\romegapi}			\\
\fomegapiz	& $-0.9^{+5.0}_{-3.2}$	& 18		& 88.8		& 		& \bma{\romegapiz} 	&($<\uomegapiz$)\\
\dbline
\end{tabular}
\end{center}
\end{table}

We compute the branching fractions from the fitted signal event yields,
reconstruction efficiency, daughter branching fractions, and the number
of produced $B$ mesons, assuming equal production rates of charged and
neutral pairs.  
To determine the reconstruction efficiency, including any yield bias of
the likelihood fit, we apply the method to simulated samples with the
signal and continuum background populations expected in the data.
Table \ref{t:results} shows for each decay chain the branching fraction
we measure, together with the quantities entering into its computation.
The statistical error on the number of events is taken as the shift from
the central value that changes the quantity $\chi^2\equiv -2\log{(\cal
L/\cal L_{\rm max})}$ 
by one unit. We also give the significance $S$, computed as the
square root of the difference between the value of $\chi^2$ for zero
signal and the value at its minimum.  The $\chi^2$ used for significance
includes a term that accounts for the additive systematic
error discussed below.  Where the significance is less
than four standard deviations, we quote also (Bayesian) 90\% C.L. upper limits,
defined by the solution $B$ to the condition $\int_0^B{\cal
L}(b)db/\int_0^\infty{\cal L}(b)db=0.9$. 

\providecommand{\psfile}[3][]{ 
  \begin{center}
    \setlength{\epsfxsize}{#3\linewidth}\leavevmode
    \def\noOpt{}\def\testit{#1}\ifx\testit\noOpt%
      \epsfbox{#2}%
    \else%
      \epsfbox[#1]{#2}%
    \fi
  \end{center} 
}

\begin{figure}[htbp]
 \psfile[72 157 532 608]{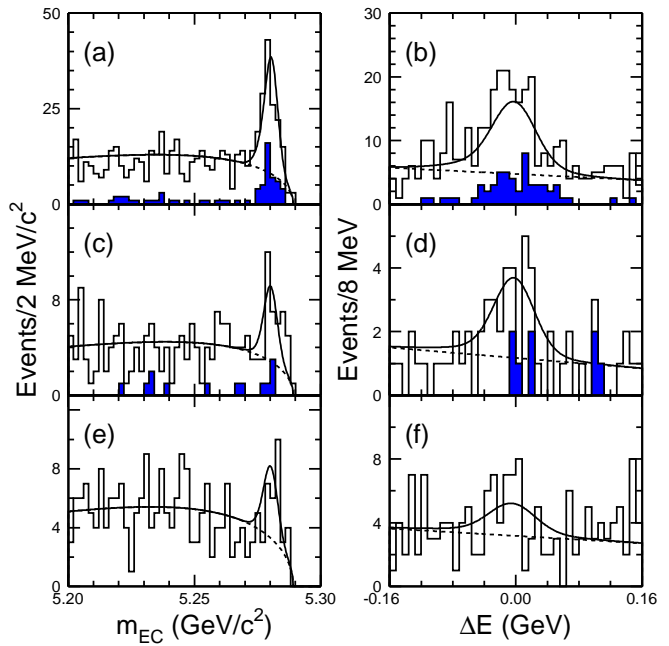}{1.0}
\vspace{-0.5cm}
 \caption{\label{fig:projMb}
$B$ candidate \mb\ and \DE\
for \etapKp (a, b), \etapKz (c, d), and
\omegapi (e, f).  Histograms represent data, with the
$\eta^\prime\ra\eta\pi\pi$ subset shaded, solid curves represent
the full fit functions, and dashed curves the background functions.  }
\end{figure}

In Fig.\ \ref{fig:projMb}\ we show projections of \mb\ and \DE\
for the modes
with significant yields.  The projections are made by selecting events
with signal likelihood (computed without the variable plotted) exceeding 
a  mode-dependent threshold that optimizes the expected sensitivity.

%
%
We have evaluated systematic errors, which are dominated in most cases
by the PDF uncertainties (3--18\%, depending
on the decay mode).  To determine these we vary parameters of the PDFs
within their uncertainties and observe the impact on the fit yield.
We include them in upper limits by
convolution with the likelihood function.
This is the only additive systematic error; all others are
multiplicative.  The estimate of any systematic bias from the fitter itself
(1--4\%) comes from fits of simulated samples with varying background
populations.

The uncertainty in our knowledge of the efficiency is found from
auxiliary studies to be 1\%\ per track, 1.25\%\ per photon, and 5\%\ per
\KS\ for the candidate $B$ and the unreconstructed $B$, which must
contribute tracks to fulfill the event multiplicity requirement.  We add
these errors linearly for the required tracks in the event, and
similarly for the photons and neutral kaons.  Our estimate of the $B$
production systematic error is 1.6\%.  Published world averages
\cite{pdg}\ provide the $B$ daughter branching fraction uncertainties.

Systematic errors associated with the event selection are minimal given
the generally loose requirements.  We account explicitly for \costhr\
(1\%), for which we observe a nearly uniform distribution in the signal
simulation.  We also include errors of 4\%\ from those PID requirements
that are imposed via cuts rather than the fit.

%
%
We have observed signals of at least 4$\sigma$ in five of the decay chains
studied here, as reported in Table \ref{t:results}.  Where we have
multiple chains for a given mode we 
combine the results by adding the $\chi^2$ distributions that represent
them and their uncorrelated statistical and systematic errors.  

The final results are
generally in agreement with those previously
reported \cite{CLEOomega,CLEOetapr}, with somewhat smaller errors.  In
particular, we confirm the expected $\Bomegapi > \BomegaKp$, and the
rather larger than predicted \cite{thy}\ rate for \etapK\ obtained by
the CLEO Collaboration \cite{CLEOetapr}.  Conjectured sources of
$\eta^\prime$ enhancement 
include flavor singlet \cite{hairpin}, charm enhanced \cite{charming},
and constructively 
interfering internal penguin diagrams \cite{thy,lipkin}.
Our results in combination with expected measurements of related modes
involving $\eta$ and $K^*$ should help to clarify this situation.

%
We are grateful for the excellent luminosity and machine conditions
provided by our \pep2\ colleagues.
The collaborating institutions wish to thank 
SLAC for its support and kind hospitality. 
This work is supported by
DOE
and NSF (USA),
NSERC (Canada),
IHEP (China),
CEA and
CNRS-IN2P3
(France),
BMBF
(Germany),
INFN (Italy),
NFR (Norway),
MIST (Russia), and
PPARC (United Kingdom). 
Individuals have received support from the Swiss NSF, 
A.~P.~Sloan Foundation, 
Research Corporation,
and Alexander von Humboldt Foundation.


\begin{thebibliography}{99}
\bibitem{bib:conjugate} 
The charge conjugate of the named state is implicitly included,
here and throughout this paper.
\end{thebibliography}
\end{document}